\documentclass[a4paper,11pt]{article}
\pdfoutput=1 

\usepackage{jcappub} 

\usepackage{graphicx}
\usepackage{amsmath}
\usepackage{amssymb}
\usepackage[all]{hypcap}
\usepackage{mathptmx}
\usepackage{txfonts}
\usepackage{ae,aecompl}
\usepackage[normalem]{ulem}
\usepackage{longtable}
\usepackage{multirow}
\usepackage{multicol}
\usepackage{tablefootnote}
\usepackage{appendix}
\usepackage{xcolor}
\usepackage{pagecolor}
\usepackage{hyperref}
\usepackage{chngcntr}
\usepackage{orcidlink}

\definecolor{hcolo}{RGB}{0,0,190}
\hypersetup{colorlinks=true,linkcolor=hcolo,citecolor=hcolo,filecolor=hcolo,urlcolor=hcolo}

\title{Galaxy evolution in the cosmic web: the relative impact of nodes and filaments in the EAGLE simulation}

\author[a]{Suman Sarkar\,\orcidlink{0000-0002-5465-3467}}
\author[b]{, Biswajit Pandey\,\orcidlink{0000-0001-7876-595X}}
\author[c]{, and Apashanka Das\,\orcidlink{0000-0002-6841-9604}}

\affiliation[a]{{\small Department of Physics, Biswa Bangla Biswabidyalay, Bolpur, West Bengal - 731204, India.}}
\affiliation[b]{{\small Department of Physics, Visva-Bharati University, Santiniketan, West Bengal - 731235, India}}
\affiliation[c]{{\small Harish-Chandra Research Institute, HBNI, Chhatnag Road, Jhunsi, Allahabad, 211109, India}}

\emailAdd{suman2reach@gmail.com}
\emailAdd{biswap@visva-bharati.ac.in}
\emailAdd{a.das.cosmo@gmail.com}

\abstract{Galaxies evolve within the intricate geometry of the cosmic web, yet the distinct physical roles of its primary components - nodes and filaments remain incompletely understood. Using the EAGLE cosmological hydrodynamical simulation, we investigate how proximity to filament spines and nodes jointly and independently regulates galaxy evolution. Galaxies are classified into red, green, and blue populations through a fully data-driven entropic thresholding method, while nodes and filaments are identified using the topological structure finder DisPerSE. We find that red galaxies preferentially inhabit regions close to both filament cores and nodes, whereas blue and green galaxies dominate the outskirts. This spatial segregation reveals two characteristic transition scales associated with filaments and nodes. The node-driven crossover occurs at $\sim 2.5~\mathrm{Mpc}$, comparable to typical virial radii of group- and cluster-sized halos, indicating that colour transformation becomes efficient once galaxies enter dynamically bound environments where tidal interactions, mergers, and gas stripping operate effectively. In contrast, the filament-driven crossover occurs at $\sim 0.75~\mathrm{Mpc}$, tracing the transverse extent of dynamically influential filament cores, where transformation proceeds more gradually through the suppression or modulation of cold gas accretion. Adopting an information-theoretic framework, we show that the normalised mutual information between dominant mass component and colour increases with distance from nodes at fixed filament proximity. This demonstrates that cluster-scale environments partially erase the intrinsic mass-colour coupling, whereas filamentary environments preserve and enhance it. The effect is strongly mass dependent: low-mass galaxies exhibit pronounced environmental modulation, while high-mass systems show weaker sensitivity, reflecting the interplay between external tidal forces and internal binding energy. Our results establish cosmic web evolution as a scale-dependent process in which nodes act as nonlinear dynamical processors, while filaments function as extended regulators of anisotropic gas accretion.}

\begin{document}
\maketitle
\flushbottom

\section{Introduction}

Galaxies form and evolve within the cosmic web, a vast network of nodes, filaments, sheets, and voids that emerges from the gravitational amplification of primordial density fluctuations \citep{gregory78, joeveer78, einasto80, zeldovich82, einasto84, bond96, bharad00, pandey05, arag10, cautun14, libeskind18}. This large-scale structure provides the fundamental environment in which galaxies assemble their mass, acquire angular momentum, and regulate their star formation activity. While galaxy evolution is known to depend strongly on internal processes and host dark matter halo mass, it has become increasingly clear that the geometry of the cosmic web itself plays an independent and physically meaningful role, encoding information that cannot be fully captured by local overdensity alone.

Early observational studies already pointed to a close connection between galaxy populations and filamentary environments. Analyses of Sloan Digital Sky Survey (SDSS) data revealed correlations between galaxy luminosity, colour, morphology, and the degree of filamentarity of the galaxy distribution \citep{pandey06}. In particular, red galaxies and ellipticals were found to be more strongly associated with dense filament intersections, while blue spirals traced the extended filamentary network. Subsequent work showed that star-forming galaxies exhibit a more filamentary distribution than passive systems \citep{pandey08}, highlighting the intertwined roles of intrinsic galaxy properties and large-scale environment. The explicit dependence of galaxy colour on cosmic web geometry has since been established with increasing clarity. 

Direct evidence for active environmental processing within filaments has emerged from studies of galaxies infalling along supercluster-scale filaments. Investigations of filaments connecting rich clusters revealed enhanced star formation at intermediate distances from cluster centres, followed by rapid suppression closer to nodes \citep{porter07, porter08}. These trends, particularly pronounced for dwarf galaxies, were attributed to galaxy-galaxy interactions and harassment in crowded infall regions, preceding gas stripping and quenching in dense cluster cores. Such results firmly established filaments as sites of significant ``pre-processing'' \citep{fujita04}, rather than passive conduits of matter.

With the development of robust filament-identification techniques and the availability of large galaxy surveys, it has become possible to quantify galaxy properties as explicit functions of distance to filament spines. Studies using SDSS data demonstrated systematic gradients in morphology, colour, and star formation rate with proximity to filaments that persist even after controlling for local density and redshift \citep{kuutma17}. Similar trends were observed at intermediate redshift using VIPERS, where massive and quiescent galaxies were found preferentially closer to filament axes, while low-mass star-forming galaxies occupied filament outskirts \citep{malavasi17}. Crucially, these signals remain significant after removing the influence of nodes, reinforcing the notion that filament geometry encodes information beyond simple overdensity.

These observational findings are supported by analyses combining surveys and hydrodynamical simulations. Using COSMOS data and the HORIZON-AGN simulation, \citep{laigle18} detected transverse stellar mass and colour gradients around filaments that persist after accounting for nodes and local density. Similarly, \citep{kraljic18} found strong mass and colour gradients with distance to filaments and nodes in both the GAMA survey and simulations, attributing these trends to anisotropic tidal fields and filament-driven assembly bias. Further evidence linking filamentary environments to gas depletion and quenching was provided by \citep{bonjean20}, who connected gradients in stellar mass, star formation rate, and quiescent fraction around filaments to the distribution of diffuse gas. Using SDSS data, \citep{pandey20} demonstrated that filaments host a higher fraction of red galaxies than sheets at fixed luminosity and local density, thereby isolating the effect of geometry from that of overdensity. Importantly, the bimodal nature of the galaxy colour distribution was found to persist across all environments, suggesting that transitions from the blue cloud to the red sequence can occur universally, but with efficiencies that depend on the surrounding large-scale structure.

Hydrodynamical simulations offer key physical insight into the mechanisms underlying these trends. Cold gas inflows along filaments can sustain star formation in specific mass regimes, while regions near nodes favour quenching through shock heating, enhanced merger activity, and gas stripping. Using the Simba simulation, \citep{bulichi24} showed that star formation is significantly suppressed within $\lesssim 100~\mathrm{kpc}$ of filaments at low redshift, accompanied by reduced $\mathrm{H\,I}$ fractions and elevated metallicities. Comparisons across different simulation suites, including EAGLE and IllustrisTNG, reveal broad qualitative agreement on the existence of filament-driven effects, while highlighting model-dependent differences in the balance between gas accretion, heating, and feedback.

Galaxy colour provides a particularly powerful tracer of evolutionary state. The well-established bimodal colour distribution separates galaxies into the blue cloud and the red sequence \citep{strateva01, blanton03, bell03, balogh04, baldry04}, with the green valley representing a transitional population undergoing quenching \citep{wyder07, salim14}. Blue galaxies typically sustain star formation through ongoing cold gas accretion, whereas red galaxies have largely exhausted or lost their gas reservoirs. Green valley systems exhibit intermediate properties, reflecting quenching pathways influenced by both internal feedback and environment \citep{schawinski14, salim14, das21, das25}. Recent observational studies have shown that, at fixed stellar mass and local density, galaxies closer to filaments are redder and more gas poor \citep{hoosain24, zarattini25}. Complementary analyses based on tidal-tensor classifications further demonstrate that the scaling relations between galaxy properties depend sensitively on cosmic web geometry \citep{nandi24, nandi25a, nandi26}. A recent analysis \citep{nandi25b} of SDSS DR18 data reveals a distinct bifurcation in galaxy evolution, where high mass galaxies are quenched in dense clusters but maintain star formation and disk-like morphologies in low-density sheets.

Despite this substantial progress, several key questions remain open. In particular, the combined influence of filaments and nodes on galaxy colour evolution and gas content has not yet been systematically quantified within a single, self-consistent framework. Moreover, many existing studies rely on fixed or subjective cuts in colour or star formation rate to define galaxy populations, potentially biasing conclusions regarding transitional systems such as the green valley. There is therefore a clear motivation for geometry-aware, data-driven approaches that explicitly account for both filamentary and nodal environments.

In this work, we address these issues using the EAGLE cosmological hydrodynamical simulation. Filaments and nodes are identified using the topological algorithm DisPerSE, which provides well-defined, scale-independent measures of distances to cosmic web features. Galaxies are classified into blue, green, and red populations using entropic thresholding, a recently proposed data-driven method that avoids subjective parameter choices \citep{pandey24}. Our primary goal is to investigate how stellar mass, gas mass, and host halo mass vary across the blue cloud, green valley, and red sequence as functions of distance to filament spines and nodes. By explicitly separating the roles of filaments and nodes, we aim to clarify how large-scale cosmic geometry regulates the balance between gas replenishment and depletion, thereby shaping galaxy colour evolution.

\section{Data and Method of Analysis}

\subsection{The EAGLE simulations}
\label{sec:data}

This work is based on the EAGLE (Evolution and Assembly of GaLaxies and their Environments) project \citep{schaye15, crain15}, a suite of state-of-the-art cosmological hydrodynamical simulations designed to model galaxy formation within the standard $\Lambda$CDM framework. The simulations self-consistently follow the coupled evolution of dark matter and baryonic matter, incorporating key physical processes such as radiative cooling, star formation, stellar evolution, chemical enrichment, and feedback from stars and active galactic nuclei. A spatially flat cosmology is adopted, consistent with \textit{Planck} results, with parameters $\Omega_{\Lambda} = 0.693$, $\Omega_{m} = 0.307$, $\Omega_{b} = 0.048$, and $H_{0} = 67.77 \, \mathrm{km\,s^{-1}\,Mpc^{-1}}$ \citep{planck2014}. The simulations evolve matter from an initial redshift of $z=127$ to the present epoch ($z=0$) within periodic, comoving cubic volumes of side lengths $25$, $50$, and $100~\mathrm{Mpc}$, using a modified version of the GADGET-2 code \citep{springel05}.

In this work, we analyse the largest EAGLE reference simulation, the $100~\mathrm{Mpc}$ comoving box \texttt{Ref-L0100N1504}, at $z=0$ (\textit{snapnum = 28}). This volume provides an optimal balance between spatial resolution and statistical power, allowing for a robust characterisation of large-scale cosmic web structures while resolving galaxy-scale baryonic processes. The simulation contains $2 \times 1504^{3} \approx 6.8$ billion particles, with an initial baryonic particle mass of $1.81\times10^{6}\,M_{\odot}$ and a dark matter particle mass of $9.70\times10^{6}\,M_{\odot}$. At this resolution, galaxies with stellar masses above $\log(M_{\star}/M_{\odot}) \sim 8$ are well resolved, enabling reliable measurements of their stellar, gaseous, and photometric properties. Galaxies are identified using the standard EAGLE group-finding pipeline, in which dark matter halos are first located using a friends-of-friends algorithm and substructures are subsequently identified with SUBFIND \citep{springel01}. The positions of galaxies are defined by the minima of their gravitational potential, as provided in the \texttt{Ref-L0100N1504\_Subhalo} catalogue \citep{mcalpine16}. To ensure that our sample contains only genuine galaxies, we retain only objects with \texttt{Spurious = 0}. The \texttt{Spurious} flag identifies artefacts of the \textsc{SUBFIND} algorithm, such as dense stellar clumps or massive black holes that may occasionally be misidentified as independent subhaloes \citep{schaye15,mcalpine16}. These objects do not correspond to physical galaxies and are therefore excluded from our analysis.

We extract a number of physical properties of the selected galaxies from the \texttt{Ref-L0100N1504\_Aperture} catalogue, measured within a spherical aperture of comoving radius $30~\mathrm{kpc}$. These include stellar mass, gas mass, dark matter mass, and the rest-frame $(u-r)$ colour. The use of a fixed physical aperture ensures consistency across the galaxy population and facilitates direct comparisons between different evolutionary classes. We restrict our analysis to galaxies with stellar masses $\log(M_{\star}/M_{\odot}) \geq 8.3$, which corresponds to the lower mass limit above which reliable $(u-r)$ colour information is available in the EAGLE database. At this threshold, each galaxy is resolved by at least $\sim 110$ particles, ensuring robust measurements of both stellar and gaseous components. Applying these selection criteria yields a final sample of $29\,754$ galaxies at $z=0$. This galaxy sample forms the basis for our analysis of how stellar mass, gas content, and halo properties vary across the blue cloud, green valley, and red sequence as functions of distance to filaments and nodes in the cosmic web. The present study focuses on the $z=0$ snapshot, which provides the most direct comparison with the large body of low-redshift observational studies investigating galaxy populations across the cosmic web. While the EAGLE simulation enables a similar analysis at earlier epochs, the strength of environmental effects is expected to evolve with cosmic time owing to changes in halo growth, gas accretion, merger rates, and the maturation of the cosmic web. A systematic investigation of this redshift evolution is beyond the scope of the present work and will be explored in a future study.

\begin{figure*}
\centering
\includegraphics[width=\columnwidth]{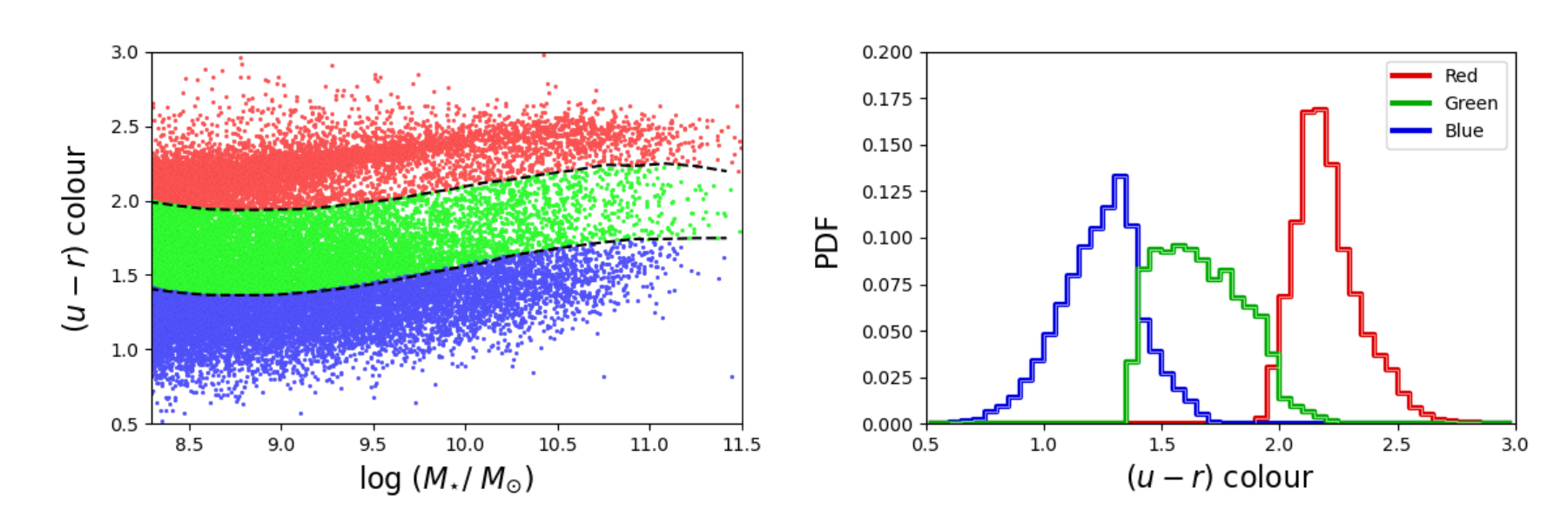}
\caption{\sf The left panel shows the red, green and blue galaxies identified using entropic thresholding on the $(u-r)$ colour-stellar mass plane. The PDFs of the three types of galaxy are shown in the right panel.}
\label{fig:classentropic}
\end{figure*}

\subsection{Classifying red, blue, and green galaxies using entropic thresholding}
\label{sec:ent_thres}

Galaxy colour provides a direct observational proxy for evolutionary state, exhibiting a well-known bimodal distribution that separates star-forming systems in the blue cloud from passive galaxies in the red sequence, with the green valley occupying the intermediate, transitional regime. In this work, we adopt a fully data-driven approach to classify galaxies into blue, green, and red populations at $z=0$, using a combination of variance-based and entropy-based thresholding techniques.

Galaxy populations are commonly classified using a variety of techniques, including fixed colour cuts \citep{strateva01}, empirical boundaries in colour-stellar mass diagrams \citep{bell04}, selections based on specific star formation rate \citep{wetzel12}, UVJ colour-colour classifications \citep{williams09}, and statistical approaches such as mixture modelling \citep{taylor15}. While these methods have been widely used, the resulting boundaries often depend on empirical choices, adopted models, or externally imposed thresholds, leading to some degree of ambiguity in the identification of transitional systems. This issue is particularly relevant for green valley galaxies, whose definition can vary substantially between studies. In this work, we adopt the entropic thresholding method proposed by \citep{pandey24}, which provides a fully data-driven classification of red, green, and blue galaxies. The method determines the optimal boundaries directly from the colour and stellar mass distributiona and therefore avoids subjective choices regarding the location of the transitions between populations.

As a first step, we identify the primary bimodality in the $(u-r)$ colour distribution using Otsu’s method, following \citep{pandey23}. This technique determines an optimal colour threshold that separates the full galaxy population into two classes by minimizing the variance within each class while simultaneously maximizing the variance between them. The two resulting populations correspond to the Blue Cloud (BC), dominated by star-forming galaxies with younger stellar populations, and the Red Sequence (RS), comprising predominantly passive galaxies with older stellar populations.

Formally, Otsu’s method minimizes the intra-class variance,
\begin{equation}
\sigma^2_{\mathrm{intra}} = P_{\mathrm{BC}} \sigma^2_{\mathrm{BC}} + P_{\mathrm{RS}} \sigma^2_{\mathrm{RS}},
\end{equation}
while maximizing the inter-class variance,
\begin{equation}
\sigma^2_{\mathrm{inter}} = P_{\mathrm{BC}} P_{\mathrm{RS}} \left(\mu_{\mathrm{BC}} - \mu_{\mathrm{RS}}\right)^2,
\end{equation}
where $P_{\mathrm{BC}}$ and $P_{\mathrm{RS}}$ denote the probabilities of occurrence of the two classes, and $\mu$ and $\sigma^2$ represent their respective mean colours and variances. This procedure yields the characteristic mean colours $\mu_{\mathrm{BC}}$ and $\mu_{\mathrm{RS}}$, which define the centres of the blue cloud and red sequence.

Otsu’s algorithm works by scanning through possible thresholds to find the one that best separates the classes. Essentially, it seeks a balance where the differences within each class are as small as possible, while the differences between classes are as large as possible. Interestingly, the threshold that achieves the lowest intra-class variance automatically produces the highest inter-class variance. This means that either $\sigma^2_{\mathrm{intra}}$ or $\sigma^2_{\mathrm{inter}}$ can be used interchangeably to identify the optimal threshold for red/blue classification.

While this binary classification robustly captures the dominant bimodality, it does not isolate the green valley population that lies between the two peaks. To explicitly identify this transitional class, we employ the entropic thresholding framework introduced by \citep{pandey24}, which is based on the principle of maximum entropy \citep{kapur85}. This method enables an objective partitioning of the colour distribution into three statistically distinct classes without imposing arbitrary colour cuts.

The colour distribution in the interval under consideration is represented by a normalized histogram consisting of $N$ bins. The colour interval between $\mu_{\mathrm{BC}}$ and $\mu_{\mathrm{RS}}$ is first discretized into $N$ bins, with $N=30$ adopted in this work. Previous studies have shown that the resulting thresholds are not sensitive to the precise choice of bin number \citep{pandey24}. Two thresholds, $s_1$ and $s_2$ (with $s_1 < s_2$), are then identified within this interval by maximizing the total entropy,
\begin{equation}
H_{\mathrm{total}} = H_{\mathrm{BC}} + H_{\mathrm{GV}} + H_{\mathrm{RS}},
\end{equation}
where the entropy associated with each class is given by
\begin{equation}
H_{\mathrm{BC}} = \log\!\left(\sum_{i=1}^{s_1} p_i\right) - 
\frac{\sum_{i=1}^{s_1} p_i \log p_i}{\sum_{i=1}^{s_1} p_i},
\end{equation}
\begin{equation}
H_{\mathrm{GV}} = \log\!\left(\sum_{i=s_1+1}^{s_2} p_i\right) - 
\frac{\sum_{i=s_1+1}^{s_2} p_i \log p_i}{\sum_{i=s_1+1}^{s_2} p_i},
\end{equation}
\begin{equation}
H_{\mathrm{RS}} = \log\!\left(\sum_{i=s_2+1}^{N} p_i\right) - 
\frac{\sum_{i=s_2+1}^{N} p_i \log p_i}{\sum_{i=s_2+1}^{N} p_i},
\end{equation}
Here $p_i$ represents the normalized histogram count (relative frequency) of galaxies in the $i^{\mathrm{th}}$ colour bin, such that $\sum_i p_i = 1$. These quantities are computed directly from the observed colour distribution and are used to evaluate the entropy of the different classes. The optimal values of $s_1$ and $s_2$ are obtained by iterating over all admissible threshold pairs and selecting the combination that maximizes $H_{\mathrm{total}}$. This procedure partitions the galaxy population into three classes: the Blue Cloud ($(u-r) \leq s_1$), the Green Valley ($s_1 < (u-r) < s_2$), and the Red Sequence ($(u-r) \geq s_2$).

Since galaxy colour is known to correlate strongly with stellar mass, we perform this classification independently within discrete stellar mass bins. Within each mass bin, Otsu’s method is first applied to identify $\mu_{\mathrm{BC}}$ and $\mu_{\mathrm{RS}}$, after which entropic thresholding is used to determine the green valley boundaries. This ensures that the colour cuts adapt dynamically with stellar mass, yielding a physically meaningful and internally consistent classification across the full mass range.

The left panel of \autoref{fig:classentropic} illustrates this classification, where the dashed curves represent smooth cubic polynomial fits to the boundaries separating the blue cloud, green valley, and red sequence in the colour-stellar mass plane. The right panel of \autoref{fig:classentropic} displays the resulting $(u-r)$ colour probability distribution functions for the blue, green, and red galaxy populations at $z=0$, demonstrating the effectiveness of the method in isolating the transitional green valley population.


\subsection{Identifying the cosmic web environments using DisPerSE}
\label{sec:disperse}

In this work, we characterise the large-scale environment of galaxies in the EAGLE simulation by quantifying their geometric relationship to the cosmic web, specifically through their distances to filament spines and nodes. To this end, we identify filaments and nodes using the DIScrete PERsistent Structures Extractor (DisPerSE; \citep{sousbie11a,sousbie11b}), a scale-independent, topology-based algorithm designed to robustly map the filamentary structure of the cosmic web.

DisPerSE identifies cosmic web features by constructing the discrete Morse-Smale complex of the underlying density field. The density field is estimated from the discrete galaxy distribution using the Delaunay Tessellation Field Estimator (DTFE; \citep{schaap00, weygaert09}), which provides an adaptive, resolution-independent reconstruction of the continuous density field. Based on this field, DisPerSE locates the critical points of the density gradient classified by their Morse index as minima, saddle points, and maxima and identifies the integral lines connecting them. In three dimensions, these critical points correspond to voids (index 0), walls (index 1), filaments (index 2), and nodes (index 3). Filament spines are traced by the Morse field lines connecting saddle points to maxima, while nodes correspond to the local maxima of the density field where multiple filaments intersect.

To ensure that only physically meaningful structures are retained and to suppress features arising from Poisson noise or sampling fluctuations, we apply a persistence threshold of $5\sigma$, retaining only those topological features whose persistence exceeds this level. Persistence quantifies the robustness of a structure by measuring the contrast between paired critical points, thereby providing a natural, scale-independent criterion for distinguishing genuine cosmic web features from noise.

For each galaxy in the final sample, we compute two geometric quantities: the distance to the nearest filament spine, denoted as $d_{\mathrm{f}}$, and the distance to the nearest node, denoted as $d_{\mathrm{n}}$. These distances are evaluated in three-dimensional comoving space and provide complementary measures of a galaxy’s location within the cosmic web. Throughout this work, we use the pair $(d_{\mathrm{f}}, d_{\mathrm{n}})$ to characterise the cosmic web environment of galaxies, enabling us to disentangle the relative and combined influences of filamentary and nodal environments on galaxy properties.

\begin{figure*}
\centering
\includegraphics[width=\textwidth]{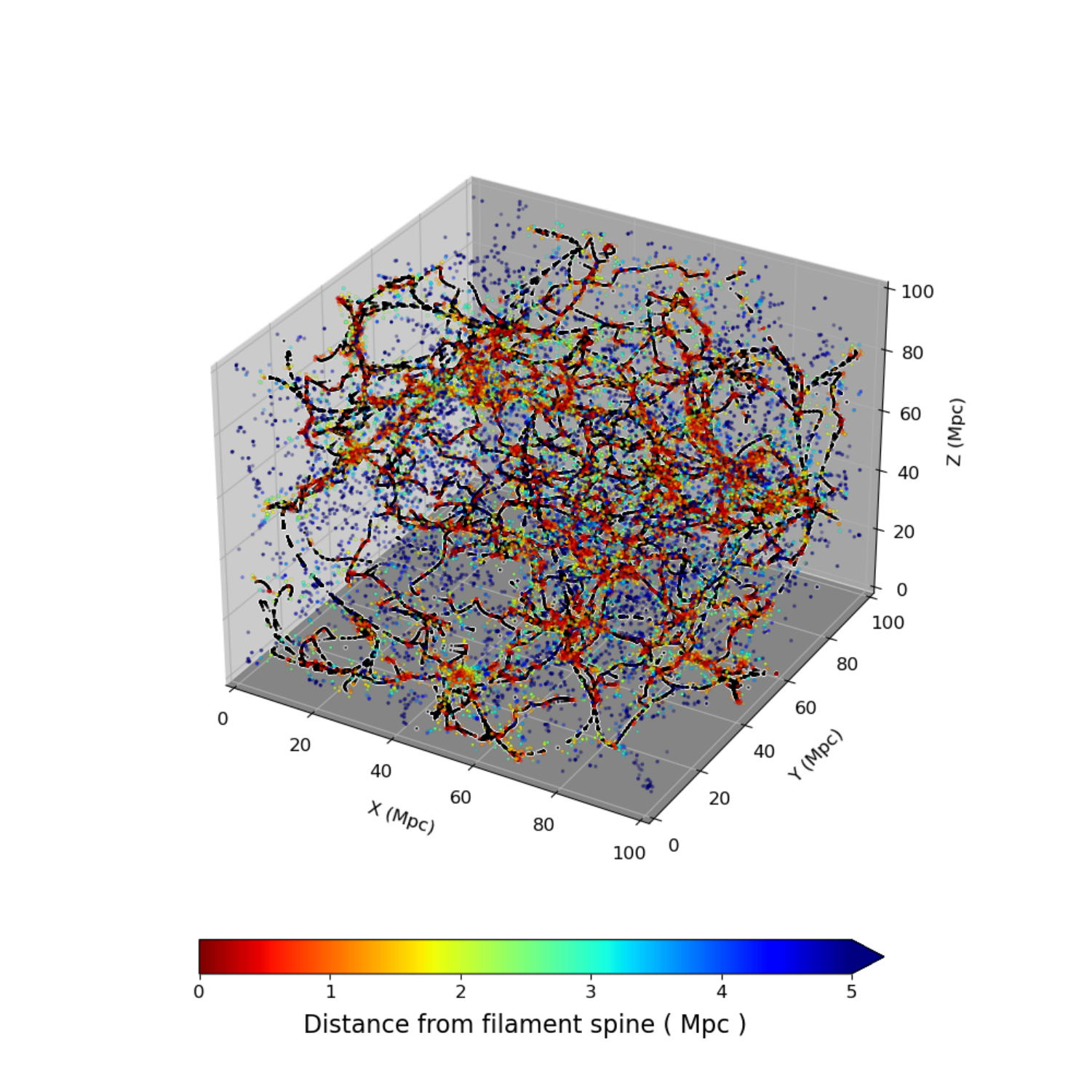}
\caption{Spatial distribution of galaxies relative to the filament spines and nodes identified by \textsc{DisPerSE}. The colour scale indicates the distance from the nearest filament spine. This figure is intended as a visual illustration of the reconstructed cosmic web geometry and the locations of galaxies with respect to its principal structural components.}
\label{fig:dist}
\end{figure*}

\subsection{Environmental trends in relative colour fractions and mass dependence}
\label{sec:frac_motivation}

To understand how galaxies migrate between the blue cloud, green valley, and red sequence, it is essential to examine how these populations are distributed across the cosmic web. While individual galaxy properties inform specific evolutionary pathways, the relative fractions of red, green, and blue galaxies provide a direct statistical measure of environmental transformation efficiency.

Relative fractions are particularly informative because absolute number counts are strongly modulated by the underlying density field, which increases toward filament spines and nodes. In contrast, quantities such as red/total or blue/total isolate changes in population composition from simple variations in galaxy abundance. This enables a cleaner assessment of how the probability of belonging to a given colour class depends on cosmic web geometry.

We therefore study the variation of red, green, and blue fractions as functions of distance to the nearest node ($d_{\mathrm{n}}$) and filament spine ($d_{\mathrm{f}}$). These two measures trace physically distinct environments: nodes correspond to dynamically evolved, virialised regions dominated by mergers, tidal heating, and hot gas, whereas filaments represent anisotropic inflow channels where gas accretion and large-scale tides regulate galaxy growth more gradually. Examining colour fractions as independent functions of $d_{\mathrm{n}}$ and $d_{\mathrm{f}}$ allows us to disentangle their relative and combined influence.

This approach is closely related to anisotropic assembly bias, in which halo growth and baryonic accretion histories depend on large-scale tidal geometry at fixed mass. To assess how such environmental effects depend on intrinsic scale, we divide the sample at the median total mass, $\log(M_{\mathrm{total}}/M_{\odot}) = 10.6$, producing statistically comparable low- and high-mass subsamples. This enables us to test whether colour transformations are primarily driven by mass, environment, or their coupled interplay across the cosmic web. \autoref{fig:dist} provides a visual representation of the spatial distribution of galaxies relative to the filament spines and nodes identified by \textsc{DisPerSE}. The figure is intended to illustrate the geometry of the reconstructed cosmic web and the locations of galaxies within it. The quantitative analyses of environmental trends and mass dependence are presented in the following sections.

\begin{table*}{}
\centering
\caption{Classification of galaxies by dominant mass component using relative mass contrast $\Delta_m$, with corresponding counts and percentages.}
\label{tab:ri}
\begin{tabular}{|l|c|c|r|}
\hline
\textbf{Galaxy class} & \textbf{Index} & \textbf{Criteria} &  \textbf{Count (percentage)} \\
\hline
Rich in stellar component & 1 & $\Delta_{star} >= \Delta_{gas} \quad \& \quad \Delta_{star} > \Delta_{DM}$ & 5786 ( \textit{24\%})\\ 
\hline
 Rich in dark matter & 2 &$\Delta_{DM} >= \Delta_{star} \quad \& \quad \Delta_{DM} > \Delta_{gas}$ & 12576 ( \textit{52\%}) \\ 
\hline
Rich in gas & 3 & $\Delta_{gas} >= \Delta_{DM} \quad \& \quad \Delta_{gas} > \Delta_{star}$ &  6021 ( \textit{25\%})\\ 
\hline
\end{tabular}
\end{table*}

\subsection{Controlling for stellar and halo mass distributions}
\label{sec:mass_control}

The environmental trends in galaxy colour fractions discussed above could in principle arise from systematic variations in the underlying stellar or halo mass distributions across the cosmic web. For example, nodes are known to host more massive halos and galaxies on average, while filament outskirts tend to be populated by lower-mass systems. Since galaxy colour correlates strongly with both stellar mass and halo mass, part of the observed environmental dependence could simply reflect this mass segregation rather than a genuine geometric effect of the cosmic web.

To assess this possibility, we repeat the relative fraction analysis within narrow bins of stellar mass and halo mass separately. By restricting the galaxy sample to a fixed and narrow mass interval, we effectively remove the first-order dependence of galaxy colour on mass. Any remaining variation of the red, green, and blue fractions with distance from filament spines ($d_{\mathrm{f}}$) or nodes ($d_{\mathrm{n}}$) can therefore be attributed primarily to environmental effects rather than differences in the mass distribution.

We compute the relative colour fractions as functions of $d_{\mathrm{f}}$ and $d_{\mathrm{n}}$ for galaxies within selected stellar mass and halo mass ranges. This approach provides a direct test of whether the characteristic environmental scales identified earlier persist when the analysis is performed at fixed mass. If the observed scale dependence were driven primarily by the variation of stellar or halo mass across the cosmic web, the environmental trends should weaken or disappear in these controlled samples. Conversely, the persistence of these trends would indicate that cosmic web geometry plays an independent role in shaping galaxy populations.

\subsection{Normalised mutual information between mass content and colour}
\label{sec:NMI}

Galaxy colour encodes evolutionary state, while the relative contributions of dark matter, stars, and gas reflect the physical processes governing galaxy assembly. A natural question therefore arises: how strongly is the dominant mass component of a galaxy linked to its colour class? 

One might attempt to address this by computing standard correlation coefficients between colour and individual mass components. However, such an approach is fundamentally limited for three reasons. First, galaxy colour in our analysis is a discrete, non-ordinal variable (red, green, blue), whereas the mass components are continuous quantities with highly non-Gaussian distributions. Second, correlations such as Pearson’s $r$ capture only linear dependencies and may fail to detect nonlinear or non-monotonic relationships. Third, the interplay between stellar, gas, and dark matter components is intrinsically multivariate and competitive where the dominance of one component necessarily suppresses the relative contribution of the others. 

To capture the full statistical dependence between colour class and mass dominance without assuming linearity or specific functional forms, we employ the framework of information theory. In particular, we quantify the association using the normalised mutual information (NMI), which measures the total shared information between two discrete random variables, irrespective of the nature of their dependence \citep{shannon48}. This approach allows us to assess whether knowledge of a galaxy’s dominant mass component reduces uncertainty about its colour class, and vice versa.

For each galaxy, the stellar mass ($M_{\mathrm{star}}$), gas mass ($M_{\mathrm{gas}}$), and dark matter mass ($M_{\mathrm{DM}}$) are known. We define the mass fraction of a component $m$ as
\begin{equation}
f_m = \frac{M_m}{M_{\mathrm{tot}}}, 
\qquad 
m \in \{\mathrm{star},\, \mathrm{gas},\, \mathrm{DM}\},
\end{equation}
where
\begin{equation}
M_{\mathrm{tot}} = M_{\mathrm{star}} + M_{\mathrm{gas}} + M_{\mathrm{DM}} .
\end{equation}

To characterise the relative dominance of each component, we introduce the mass contrast
\begin{equation}
\Delta_m = \frac{f_m}{\langle f_m \rangle} - 1 ,
\end{equation}
where $\langle f_m \rangle$ is the sample-averaged mass fraction of component $m$. Each galaxy is then assigned a discrete mass-component index according to the dominant $\Delta_m$, following the criteria summarised in \autoref{tab:ri}. This procedure partitions the galaxy population into three physically interpretable classes: stellar-dominated, dark matter-dominated, and gas-dominated systems.

Independently, galaxies are classified into red, green, and blue populations using the entropic thresholding technique described in \autoref{sec:ent_thres}. Consequently, each galaxy is characterised by two categorical labels: one describing its dominant mass component and the other describing its colour class.

Consider a sample of $N$ galaxies. Let $n_j$ denote the number of galaxies in the $j^{\mathrm{th}}$ mass-component class and $n_k$ the number in the $k^{\mathrm{th}}$ colour class, such that
\begin{equation}
\sum_{j=1}^{3} n_j = \sum_{k=1}^{3} n_k = N .
\end{equation}
Let $n_{jk}$ be the number of galaxies simultaneously belonging to mass-component class $j$ and colour class $k$. These counts satisfy
\begin{equation}
\sum_k n_{jk} = n_j, 
\qquad 
\sum_j n_{jk} = n_k, 
\qquad 
\sum_j \sum_k n_{jk} = N .
\end{equation}

The corresponding probabilities are
\begin{equation}
p(m_j) = \frac{n_j}{N}, 
\qquad 
p(c_k) = \frac{n_k}{N}, 
\qquad 
p(m_j, c_k) = \frac{n_{jk}}{N}.
\end{equation}

The Shannon entropies associated with the mass-component classification, colour classification, and their joint distribution are defined as
\begin{align}
H(M) &= -\sum_{j=1}^{3} p(m_j)\, \log p(m_j), \\
H(C) &= -\sum_{k=1}^{3} p(c_k)\, \log p(c_k), \\
H(M,C) &= -\sum_{j=1}^{3} \sum_{k=1}^{3} p(m_j,c_k)\, \log p(m_j,c_k).
\end{align}

The mutual information between mass component and colour is
\begin{equation}
I(M;C) = H(M) + H(C) - H(M,C),
\end{equation}
which quantifies the reduction in uncertainty in one variable given knowledge of the other. To facilitate comparison and interpretability, we normalise this quantity by the maximum possible entropy for three equally populated classes,
\begin{equation}
H_{\max} = \log 3,
\end{equation}
and define the normalised mutual information as
\begin{equation}
I_{\mathrm{N}}(M;C) = \frac{I(M;C)}{H_{\max}}.
\end{equation}

By construction, $0 \leq I_{\mathrm{N}} \leq 1$, where $I_{\mathrm{N}} = 0$ indicates statistical independence between dominant mass component and colour class, and $I_{\mathrm{N}} = 1$ indicates perfect association.

The normalised mutual information provides a quantitative measure of the statistical association between the dominant mass component and galaxy colour. In the present context, it can be interpreted as a measure of how strongly knowledge of a galaxy's dominant mass component (stellar-, gas-, or dark matter-dominated) constrains the probability that the galaxy belongs to the red sequence, green valley, or blue cloud, and vice versa. A larger value of $I_{\mathrm{N}}$ indicates a stronger correspondence between mass composition and colour, whereas a smaller value implies that the two classifications are more weakly related. Thus, rather than focusing on individual colour fractions, the mutual information quantifies the overall strength of the mass-colour connection within a given environment.

This information-theoretic measure provides a model-independent and non-parametric quantification of the coupling between galaxy mass composition and evolutionary state. Unlike linear correlation analyses, it captures the full structure of statistical dependence, including nonlinear, asymmetric, and multivariate effects. It therefore offers a more physically meaningful assessment of whether variations in dark matter, stellar, or gas dominance are genuinely linked to the transition between blue, green, and red galaxy populations.

\section{Results}

We now present the main findings of this work, focusing on how galaxy populations vary across the cosmic web and how the coupling between mass composition and colour depends on environment. We first examine the relative fractions of red, green, and blue galaxies as functions of distance from filament spines and nodes, and then quantify the environmental modulation of the mass-colour connection using an information-theoretic framework.

\begin{figure*}
\centering
\includegraphics[width=\textwidth]{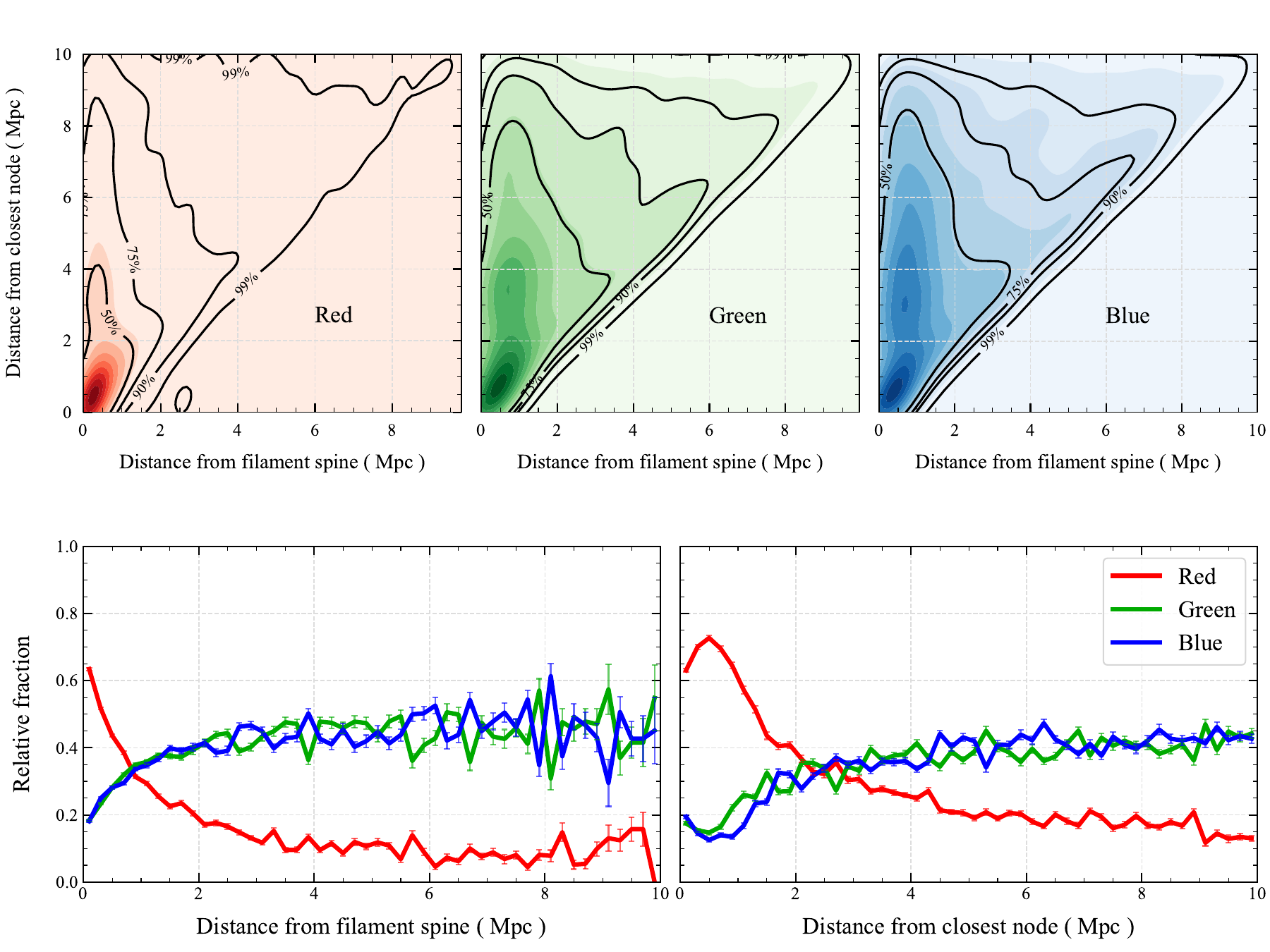}
\caption{Top panels of this figure show normalized joint distributions
  of red (left), green (middle), and blue (right) galaxies in the
  $d_{\mathrm{f}}$-$d_{\mathrm{n}}$ plane, where $d_{\mathrm{f}}$ and
  $d_{\mathrm{n}}$ denote distances from the filament spine and
  closest node, respectively. Contours enclose $50\%$, $75\%$, $90\%$,
  and $99\%$ of each population.  The bottom two panels show
  corresponding relative fractions as functions of $d_{\mathrm{f}}$
  (left) and $d_{\mathrm{n}}$ (right). Red galaxies dominate near
  filament spines and nodes, while blue galaxies preferentially occupy
  larger distances. The crossover between red and blue fractions
  occurs at $d_{\mathrm{f}} \sim 0.75~\mathrm{Mpc}$ and
  $d_{\mathrm{n}} \sim 2.5~\mathrm{Mpc}$ for filament and node
  respectively. The 1$\sigma$ error bars in each case are estimated
  using $100$ jackknife samples drawn from the original
  distributions.}
\label{fig:fig1}
\end{figure*}

\begin{figure*}
\centering
\includegraphics[width=\textwidth]{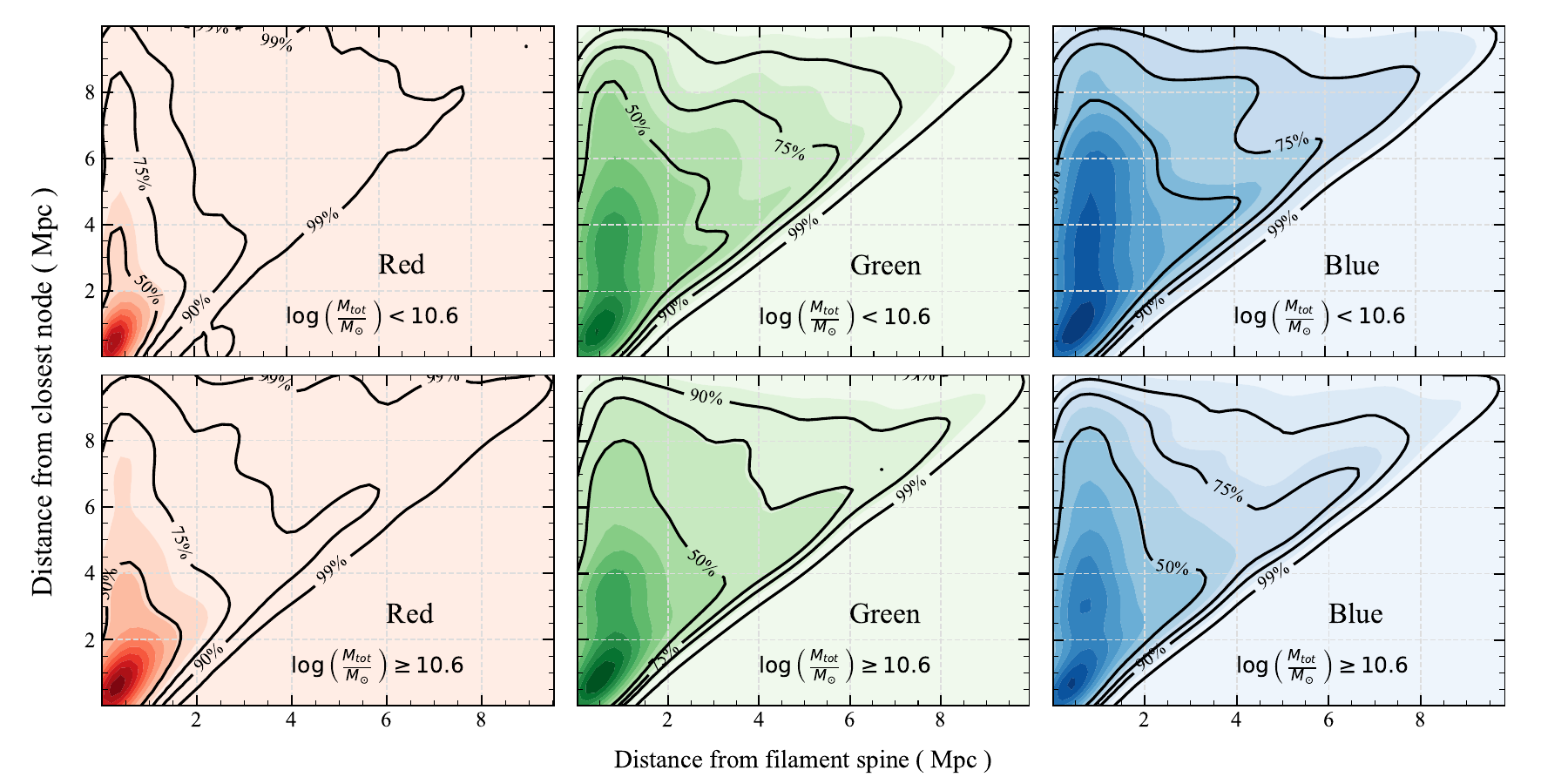}
\caption{Normalized joint distributions of red, green, and blue
  galaxies in the $d_{\mathrm{f}}$-$d_{\mathrm{n}}$ plane, shown
  separately for $\log(M_{\mathrm{tot}}/M_{\odot}) < 10.6$ (top) and
  $\log(M_{\mathrm{tot}}/M_{\odot}) \geq 10.6$ (bottom).  Low-mass
  galaxies exhibit stronger colour segregation with respect to
  filament and node proximity, whereas high-mass galaxies display a
  broader and less contrasted distribution.}
\label{fig:fig2}
\end{figure*}

\begin{figure*}
\centering
\includegraphics[width=\textwidth]{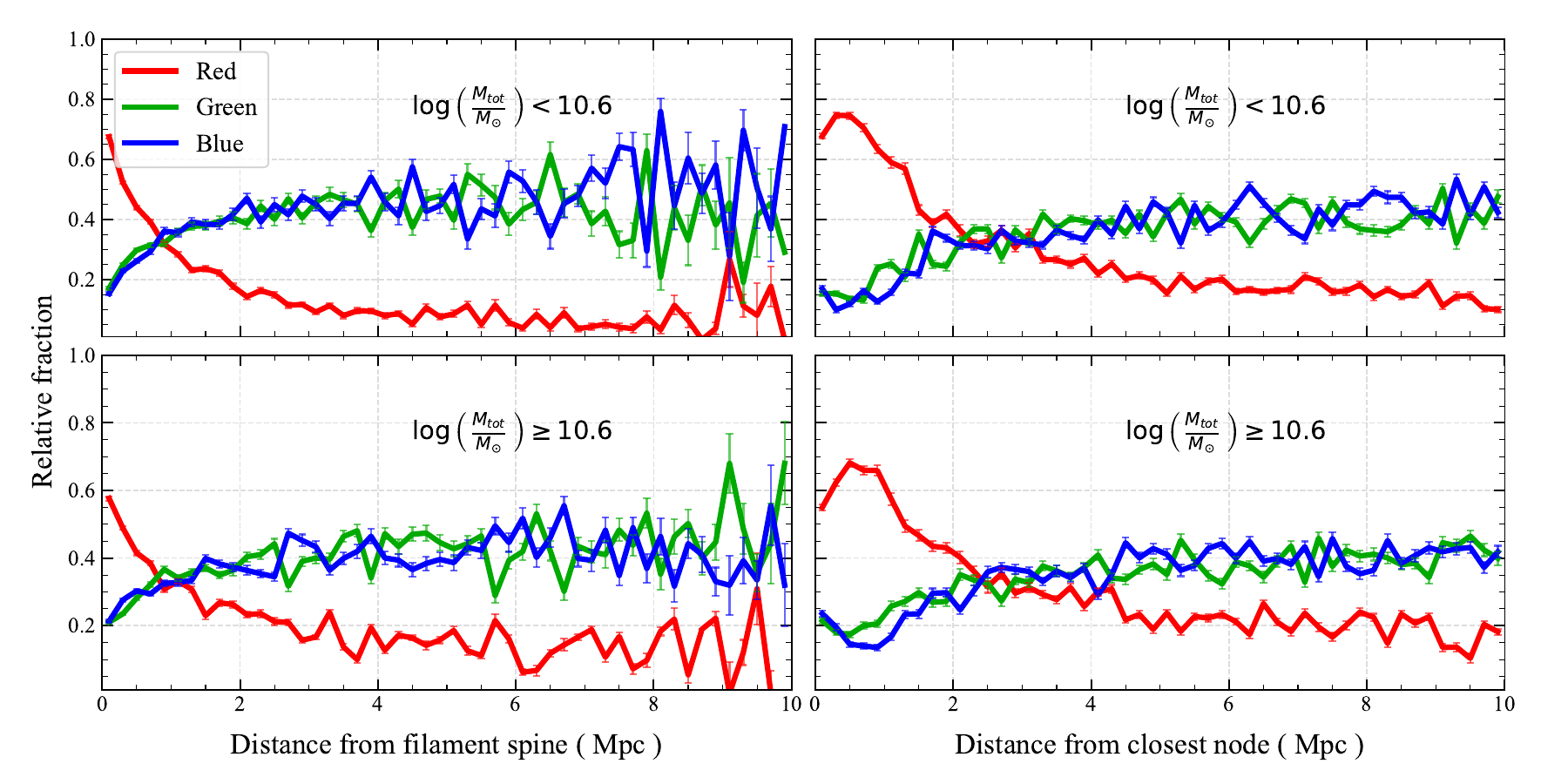}
\caption{Relative fractions of red, green, and blue galaxies as
  functions of distance from the filament spine (left) and closest
  node (right), shown separately for $\log(M_{\mathrm{tot}}/M_{\odot})
  < 10.6$ (top) and $\log(M_{\mathrm{tot}}/M_{\odot}) \geq 10.6$
  (bottom). The 1$\sigma$ errorbars are computed based on 100
  jackknife resamplings of the underlying distributions.}
\label{fig:fig3}
\end{figure*}

\begin{figure*}
\centering
\includegraphics[width=\textwidth]{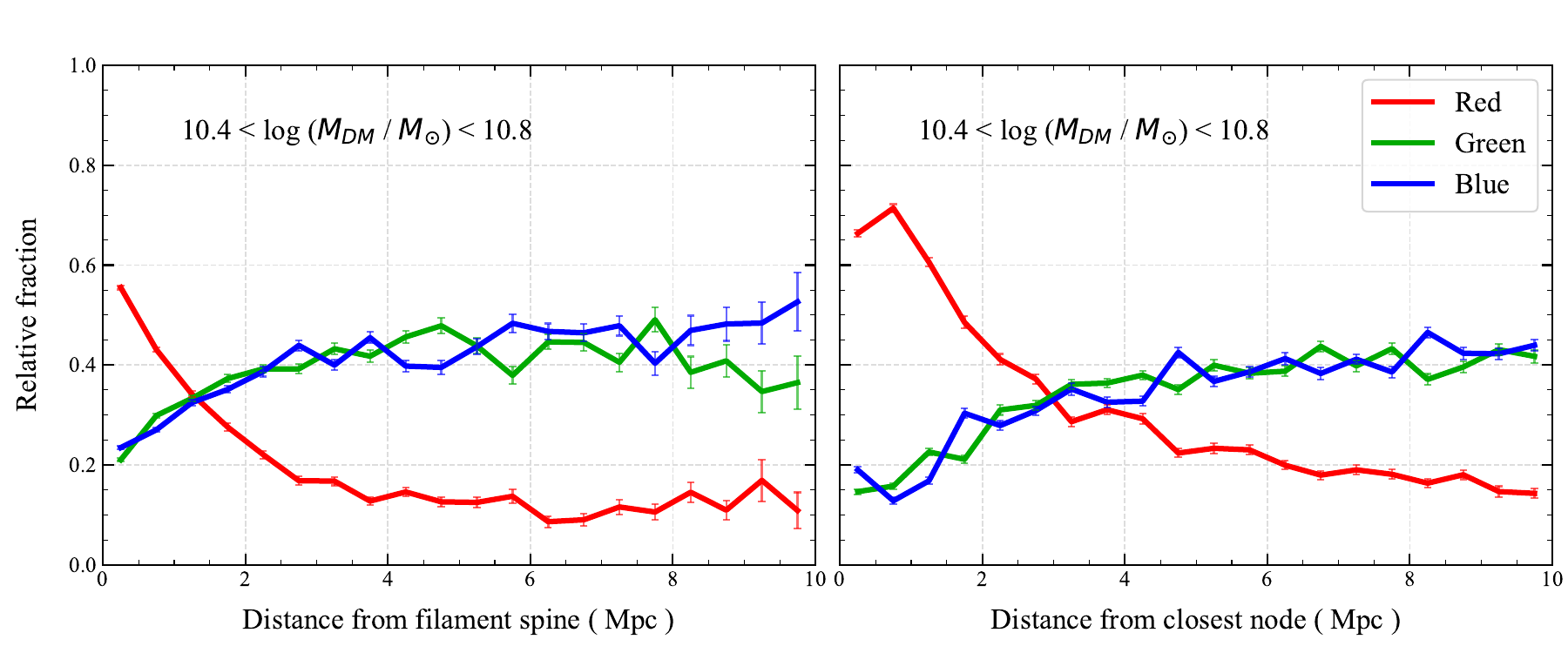}
\caption{Relative fractions of red, green, and blue galaxies as
  functions of distance from the filament spine (left) and the
  closest node (right) for galaxies within a narrow halo mass
  range $10.4 < \log(M_{\mathrm{DM}}/M_{\odot}) < 10.8$.  Despite
  restricting the sample to a fixed halo-mass interval, clear
  environmental trends remain: the red fraction decreases with
  increasing distance from nodes and filament spines, while the blue
  fraction increases correspondingly.  This demonstrates that the
  scale dependence observed in the full sample is not solely driven by
  variations in halo mass across the cosmic web.  Error bars denote
  $1\sigma$ uncertainties estimated using jackknife resampling. }
\label{fig:fixed_hm}
\end{figure*}

\begin{figure*}
\centering
\includegraphics[width=\textwidth]{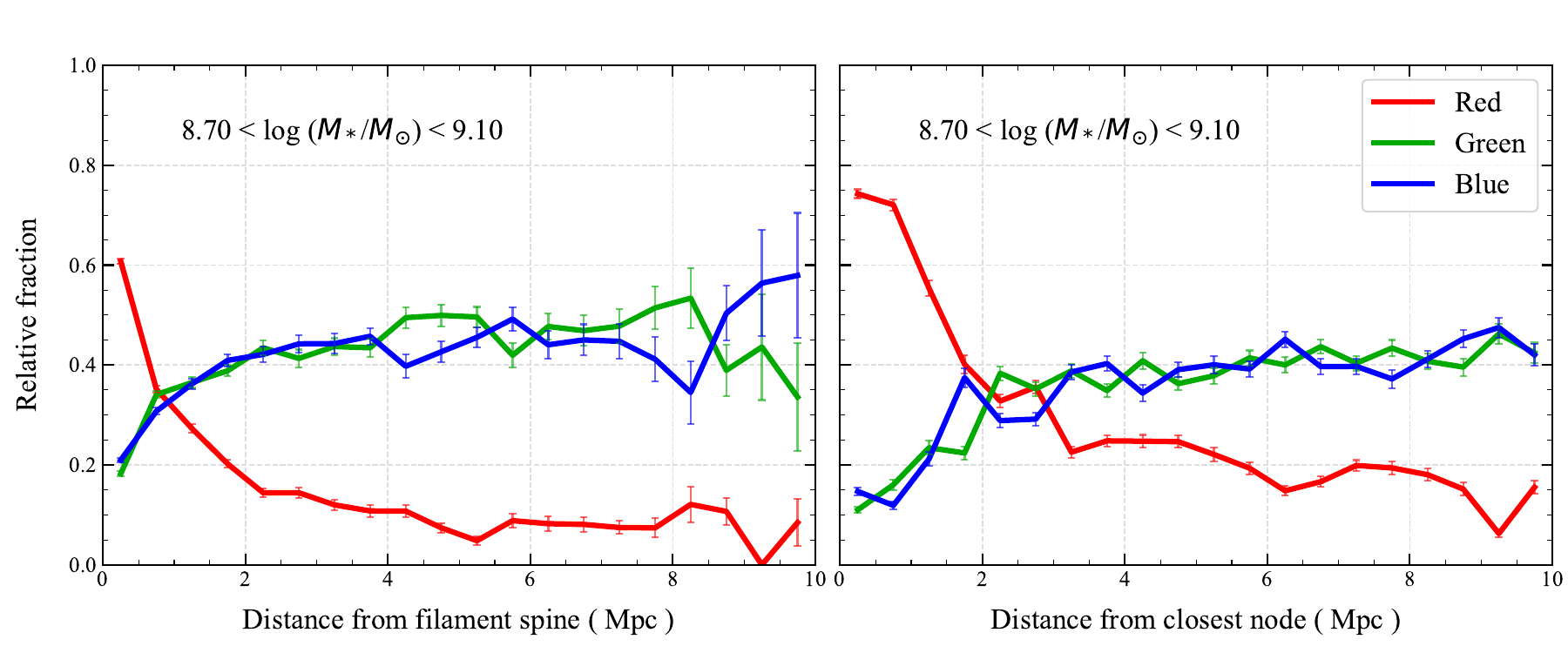}
\caption{Relative fractions of red, green, and blue galaxies as
  functions of distance from the filament spine (left) and the closest
  node (right) for galaxies within a narrow stellar mass range $8.70 <
  \log(M_{\star}/M_{\odot}) < 9.10$.  The environmental gradients
  persist even at fixed stellar mass, with red galaxies preferentially
  found near nodes and filament cores and blue galaxies becoming more
  prevalent at larger distances. While the precise transition scales
  vary slightly with the chosen mass bin, the overall behaviour
  remains unchanged, indicating that the observed scale dependence
  reflects genuine environmental effects rather than differences in
  the stellar mass distribution.}
\label{fig:fixed_sm}
\end{figure*}

\subsection{Relative fractions of red, green, and blue galaxies in the $d_{\mathrm{f}}$-$d_{\mathrm{n}}$ plane}
\label{sec:results_fractions}

\autoref{fig:fig1} presents the environmental distribution of galaxy colour classes in the two-dimensional space defined by the distance to the filament spine ($d_{\mathrm{f}}$) and the distance to the nearest node ($d_{\mathrm{n}}$). The top panels show the normalized joint distributions of red-sequence, green-valley, and blue-cloud galaxies in the $d_{\mathrm{f}}$-$d_{\mathrm{n}}$ plane, while the bottom panels display the corresponding relative fractions as one-dimensional functions of $d_{\mathrm{f}}$ and $d_{\mathrm{n}}$.

The joint distributions reveal a clear geometric segregation of galaxy populations across the cosmic web. Red galaxies are strongly concentrated toward the lower-left region of the $d_{\mathrm{f}}$-$d_{\mathrm{n}}$ plane, i.e., close to both filament spines and nodes. The empirical distribution peaks sharply at small $d_{\mathrm{f}}$ and small $d_{\mathrm{n}}$, indicating that the densest and most dynamically evolved regions of the web preferentially host passive systems. In contrast, blue galaxies occupy a larger region of the parameter space. This behaviour is consistent with blue galaxies residing in relatively low-density, dynamically quiescent environments where gas accretion remains efficient. The green-valley population occupies a region of the $d_{\mathrm{f}}$-$d_{\mathrm{n}}$ plane that is broadly consistent with its transitional nature, although its distribution shows considerable overlap with the blue-cloud population. The spatial distribution of green-valley galaxies is consistent with the environmental trends produced by the EAGLE simulation and reflects the gradual transition between the blue and red populations encoded by the physical processes operating within the model.

The bottom left panel of \autoref{fig:fig1} shows the relative fractions of red, green, and blue galaxies as functions of $d_{\mathrm{f}}$. A strong and monotonic environmental trend is evident. The red fraction decreases sharply with increasing distance from the filament spine, dropping from $\sim 0.6$ at $d_{\mathrm{f}} \lesssim 1~\mathrm{Mpc}$ to $\lesssim 0.1$ beyond $d_{\mathrm{f}} \sim 6$–$8~\mathrm{Mpc}$. Conversely, the blue fraction rises steadily with increasing $d_{\mathrm{f}}$, eventually dominating the population at $\gtrsim 0.75~\mathrm{Mpc} $. Similarly, the green fraction peaks at intermediate distances and remains comparatively stable over a broad range of $d_{\mathrm{f}}$. This behaviour demonstrates that proximity to filament spines is strongly associated with quenching. The monotonic decline of the red fraction with $d_f$ indicates that the physical processes operating within filamentary environments in the EAGLE simulation give rise to a systematic environmental dependence of galaxy populations, consistent with the modulation of gas accretion and merger histories within filament cores.

The bottom right panel  of \autoref{fig:fig1} shows the variation of relative fractions with $d_{\mathrm{n}}$. The red fraction is highest at small node distances, reaching $\sim 0.7$ at $d_{\mathrm{n}} \lesssim 1~\mathrm{Mpc}$, and decreases progressively with increasing $d_{\mathrm{n}}$. In contrast, the blue fraction increases steadily with node distance. The crossover between red and blue populations occurs at $d_{\mathrm{n}} \sim 2$–$3~\mathrm{Mpc}$, marking a characteristic transition scale between node-dominated and filament-dominated regimes. Compared to the filament trends, the node dependence appears steeper at small $d_{\mathrm{n}}$, consistent with rapid environmental processing within virialised regions. These results reveal distinct environmental scales associated with nodes and filaments. We defer a detailed physical interpretation of these scales to Section~\ref{sec:conclusions}.

These results establish a baseline environmental dependence for the full galaxy sample. In the following subsections, we examine whether these trends persist or change when the population is separated into low- and high-mass subsamples, thereby probing the interplay between intrinsic mass scale and cosmic web geometry in driving galaxy evolution.

\subsection{Mass dependence of the relative fractions in the $d_{\mathrm{f}}$-$d_{\mathrm{n}}$ plane}

To disentangle the interplay between intrinsic mass scale and large-scale environment, we also examine the same statistics separately for low-mass ($\log(M_{\mathrm{tot}}/M_{\odot}) < 10.6$) and high-mass  ($\log(M_{\mathrm{tot}}/M_{\odot}) \geq 10.6$) galaxies. \autoref{fig:fig2} and \autoref{fig:fig3} present the corresponding two-dimensional distributions and relative fractions for these subsamples.

The two-dimensional $d_{\mathrm{f}}$-$d_{\mathrm{n}}$ distributions in \autoref{fig:fig2} reveal that the geometric segregation identified in \autoref{fig:fig1} persists in both mass regimes, but with notable differences in strength.  For low-mass galaxies, the red population remains relatively more concentrated toward small $d_{\mathrm{f}}$ and small $d_{\mathrm{n}}$, while the existence of blue galaxies extend to larger distances from both structures.  The contrast between red- and blue-dominated regions is particularly sharp, indicating that low-mass systems respond strongly to cosmic web geometry. For high-mass galaxies, the segregation is somewhat weaker. Red systems are more broadly distributed across the $d_{\mathrm{f}}$-$d_{\mathrm{n}}$ plane, and blue galaxies are less spread out across the outskirts. This broader spread suggests that colour in massive galaxies is less exclusively controlled by their location in the cosmic web and more influenced by internal processes.

These differences become clearer in the one-dimensional trends observed in \autoref{fig:fig3}. For low-mass galaxies, the red fraction declines steeply with increasing distance from both filament spines and nodes. Near nodes ($d_{\mathrm{n}} \lesssim 1~\mathrm{Mpc}$), red galaxies dominate, while blue galaxies rapidly take over beyond $d_{\mathrm{n}} \sim 2.5~\mathrm{Mpc}$. Similarly, the red fraction decreases monotonically with increasing $d_{\mathrm{f}}$, and the crossover between red and blue populations occurs at comparable filament distances as in the full sample. The environmental modulation is therefore strong and systematic in this mass regime. For high mass population, the red galaxies are still more common near nodes and filament spines. However the red fraction becomes less pronounced overall. This indicates that, for massive systems, colour transformation is less tightly coupled to cosmic web position.

It may be noted that in contrast to nodes, the red fraction increases monotonically with decreasing distance to filament spines for all mass ranges considered. No turnover is observed at small $d_{\mathrm{n}}$, and the crossover between red and blue populations consistently occurs at $\sim 0.75~\mathrm{Mpc}$, independent of total mass. This behaviour suggests that filaments act as sites of gradual, large-scale environmental modulation rather than abrupt dynamical transformation. Galaxies closer to filament spines reside in regions of enhanced matter inflow, anisotropic tidal forces, and elevated merger probabilities. The steady increase in red fraction toward filament centres is consistent with a scenario of ``cosmic web starvation'', where gas accretion is progressively reduced or heated due to filamentary shocks and interactions, leading to a gradual suppression of star formation. Unlike nodes, filaments lack extremely dense cores where tidal destruction or rapid merging would significantly alter the surviving population statistics. Consequently, the red fraction does not exhibit a central turnover.

Taken together, \autoref{fig:fig2} and \autoref{fig:fig3} demonstrate that environmental colour segregation is strongly mass dependent. Low-mass galaxies exhibit pronounced gradients with both $d_{\mathrm{f}}$ and $d_{\mathrm{n}}$, consistent with their greater susceptibility to external processes such as gas stripping, tidal heating, and suppression of accretion. High-mass galaxies, possessing deeper gravitational potentials and more efficient internal feedback, show weaker environmental gradients, implying that internal quenching mechanisms compete with or dominate over large-scale environmental effects.

These results suggest that cosmic web quenching operates most efficiently in low-mass systems, where environmental modulation can significantly alter the population mix across megaparsec scales. For massive galaxies, the imprint of large-scale geometry remains visible but is partially diluted by mass-driven evolutionary pathways. The combined analysis of \autoref{fig:fig1}-\autoref{fig:fig3} therefore reveals a coherent picture in which the impact of filaments and nodes on galaxy colour is both scale dependent and intrinsically linked to galaxy mass.

\begin{figure}
\centering
\includegraphics[width= 1.5 \columnwidth/2]{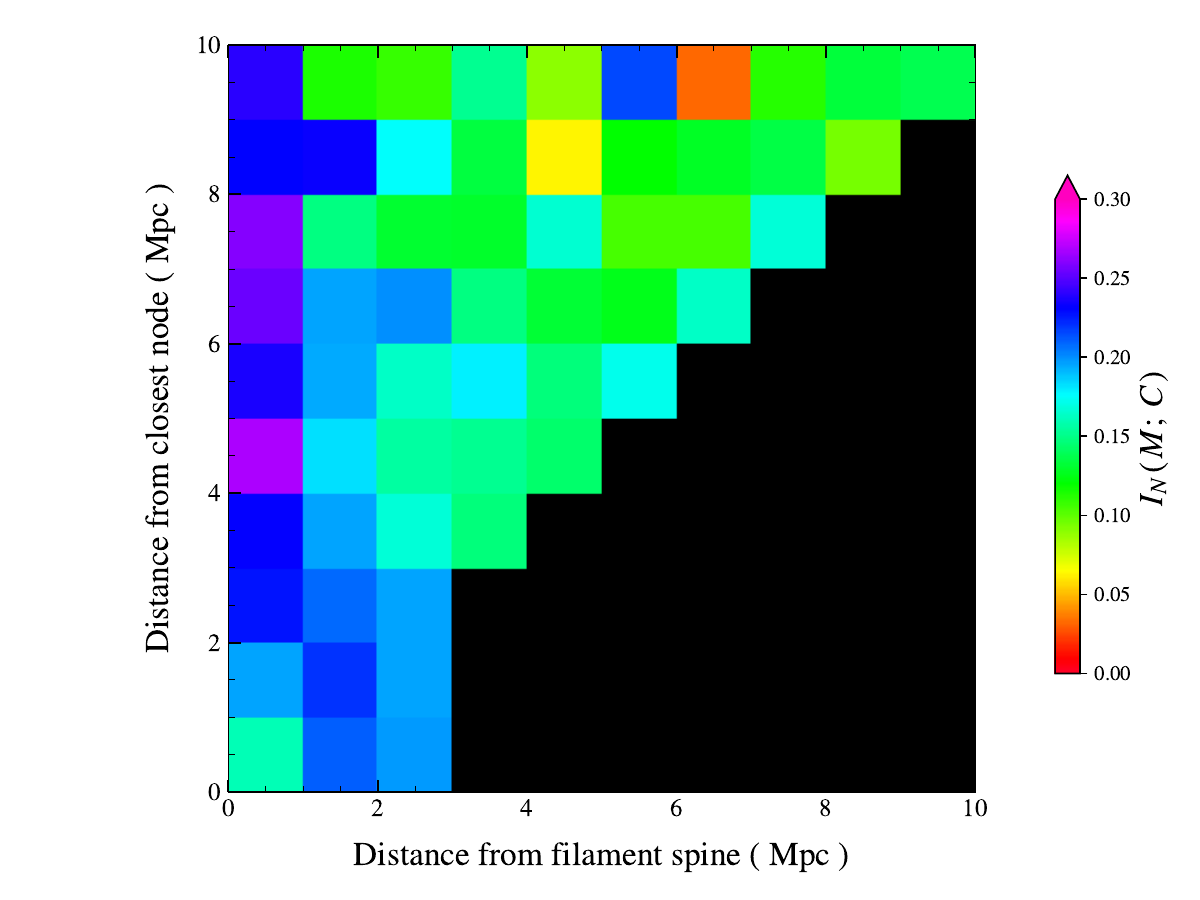}
\caption{This shows the Normalised mutual information
  $I_{\mathrm{N}}(M;C)$ between dominant mass component and galaxy
  colour across the $d_{\mathrm{f}}$-$d_{\mathrm{n}}$ plane.  Higher values indicate
  stronger statistical coupling.  The black region corresponds to
  geometrically inaccessible combinations of $(d_{\mathrm{f}},d_{\mathrm{n}})$ where no
  galaxies are present. The coupling strengthens in
  filament-dominated regions far from nodes and weakens near nodes.}
\label{fig:fig4}
\end{figure}

\begin{figure}
\centering
\includegraphics[width=1.1 \columnwidth]{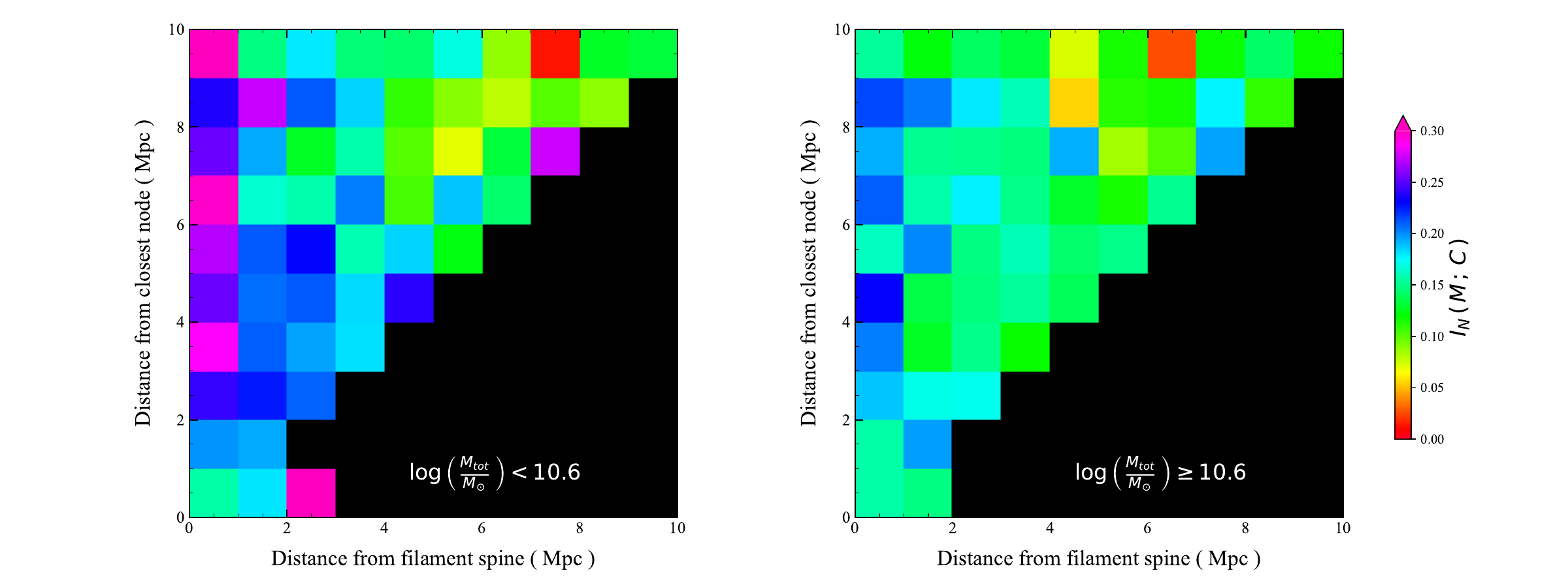}
\caption{This shows the Normalised mutual information
  $I_{\mathrm{N}}(M;C)$ in the $d_{\mathrm{f}}$-$d_{\mathrm{n}}$ plane for low-mass
  galaxies ($\log(M_{\mathrm{tot}}/M_{\odot}) < 10.6$, left) and
  high-mass galaxies ($\log(M_{\mathrm{tot}}/M_{\odot}) \geq 10.6$,
  right). The environmental modulation is stronger for low-mass
  systems, while high-mass galaxies exhibit a flatter dependence on
  cosmic web geometry.}
\label{fig:fig5}
\end{figure}

\begin{figure}
\centering
\includegraphics[width=\columnwidth]{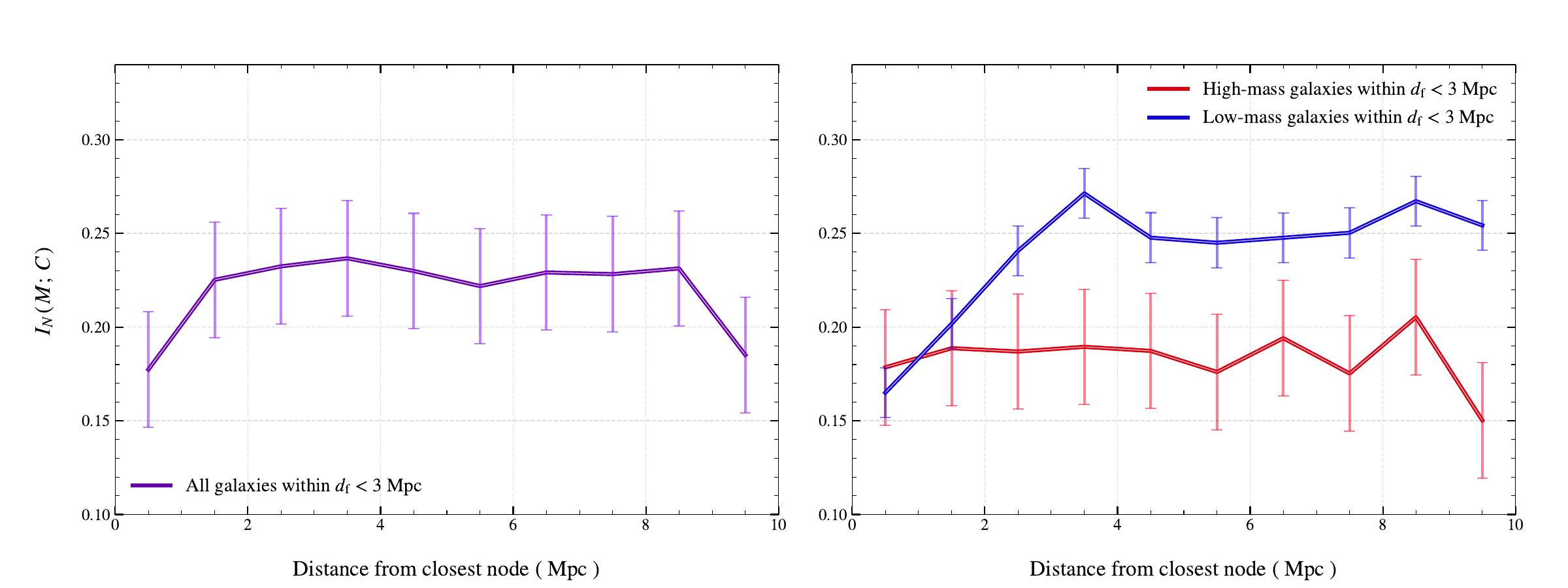}
\caption{Normalised mutual information $I_{\mathrm{N}}(M;C)$ as a
  function of node distance $d_{\mathrm{n}}$ for galaxies within
  $d_{\mathrm{f}} < 3~\mathrm{Mpc}$.  Left panel show the result for
  full galaxy sample and right panel shows the comparison between
  low-mass ($\log(M_{\mathrm{tot}}/M_{\odot}) < 10.6$) and high-mass
  ($\log(M_{\mathrm{tot}}/M_{\odot}) \geq 10.6$) galaxies. Error bars
  represent $1\sigma$ uncertainties estimated from 100 jackknife
  resamplings. Larger values of $I_{\mathrm{N}}$
    indicate a stronger statistical correspondence between the
    dominant mass component and galaxy colour. The lower
    $I_{\mathrm{N}}$ at smaller $d_{\mathrm{n}}$ highlights the
    weakening of the mass-colour coupling in the vicinity of nodes
    particularly for low-mass systems.}
\label{fig:nmi_1d}
\end{figure}

\subsection{Environmental trends at fixed stellar and halo mass}
\label{sec:fixed_mass_results}

To determine whether the environmental trends reported above are driven by variations in stellar or halo mass across the cosmic web, we repeat the analysis within narrow stellar mass and halo mass intervals. \autoref{fig:fixed_hm} and \autoref{fig:fixed_sm} show the relative fractions of red, green, and blue galaxies as functions of $d_\mathrm{f}$ and $d_\mathrm{n}$ for representative fixed halo-mass and stellar-mass bins, respectively. The stellar-mass and halo-mass intervals are centred around the respective median masses of the full sample. This choice avoids biases associated with extreme mass ranges while maintaining adequate statistics, enabling a clean test of whether the observed environmental trends persist independently of the underlying mass distribution.

The results demonstrate that the scale-dependent environmental trends persist even when the galaxy sample is restricted to narrow mass ranges. In the fixed halo-mass sample ($10.4 < \log(M_{\mathrm{DM}}/M_{\odot}) < 10.8$), the relative fractions continue to vary systematically with both filament and node distance, with red galaxies dominating near nodes and filament cores while blue galaxies become increasingly prevalent at larger separations (\autoref{fig:fixed_hm}). A similar behaviour is observed when the analysis is performed at fixed stellar mass ($8.70 < \log(M_{\star}/M_{\odot}) < 9.10$), where the relative fractions exhibit clear environmental gradients with both $d_\mathrm{f}$ and $d_\mathrm{f}$ (\autoref{fig:fixed_sm}).

Although the precise transition scales show some sensitivity to the choice of mass bin, the qualitative behaviour remains unchanged: red fractions decrease with increasing distance from nodes and filament spines, while blue fractions increase correspondingly. This persistence of the environmental trends indicates that the scale dependence identified earlier cannot be explained solely by variations in stellar or halo mass across the cosmic web.

\subsection{Environmental modulation of the mass-colour coupling using normalized mutual information (NMI)}
\label{sec:results_nmi}

We now quantify how the statistical coupling between dominant mass component and galaxy colour varies across the cosmic web using the normalised mutual information $I_{\mathrm{N}}(M;C)$.  \autoref{fig:fig4} shows the spatial distribution of $I_{\mathrm{N}}$ across the $d_{\mathrm{f}}$-$d_{\mathrm{n}}$ plane for the full galaxy sample. A clear and systematic environmental trend is evident. At small filament distances ($d_{\mathrm{f}} \lesssim 2~\mathrm{Mpc}$), $I_{\mathrm{N}}$ increases steadily with increasing node distance. For example, at a fixed $d_{\mathrm{f}} = 1~\mathrm{Mpc}$, the normalised mutual information rises from $I_{\mathrm{N}} \approx 0.15$ at $d_{\mathrm{n}} \sim 1~\mathrm{Mpc}$ to $I_{\mathrm{N}} \approx 0.28$ at $d_{\mathrm{n}} \sim 8~\mathrm{Mpc}$. In practical terms, this indicates that the dominant mass component becomes a progressively better predictor of whether a galaxy is red, green, or blue as one moves away from nodes while remaining embedded within the filament environment. In other words, this systematic rise shows that the statistical coupling between a galaxy's internal mass composition and its colour becomes increasingly pronounced in regions farther from nodes.

In the EAGLE simulation,  galaxies in the vicinity of nodes experience intense dynamical processing such as tidal interactions, ram-pressure stripping, and enhanced merger activity which can drive colour transformation independently of the galaxy’s initial baryonic configuration. Such externally driven quenching partially erases the intrinsic link between mass composition and star formation activity, thereby lowering $I_{\mathrm{N}}$. At larger node distances, where cluster-scale tidal fields weaken and filamentary regulation dominates, galaxy evolution proceeds more gradually through modulation of gas accretion. In this regime, gas-rich systems preferentially remain blue while gas-poor or stellar-dominated systems migrate toward the red sequence, restoring a tighter correspondence between internal structure and colour. 

The $I_{\mathrm{N}}$ map further reveals that the highest values ($I_{\mathrm{N}} \gtrsim 0.25$) occur in regions that are close to filaments but far from nodes. This increase in $I_{\mathrm{N}}$ demonstrates that the environmental signatures associated with filamentary regions preserve a stronger connection between mass composition and colour than those associated with nodal environments. At large $d_{\mathrm{f}}$, where both filamentary influence and gas regulation weaken, $I_{\mathrm{N}}$ decreases again, suggesting that the structural imprint of baryonic composition on colour is strongest within the dynamically influential filament core.

\subsection{Mass dependence of the NMI signal}
\label{sec:results_nmi_mass}

\autoref{fig:fig5} shows the corresponding $I_{\mathrm{N}}$ distributions after dividing the sample at the median total mass. The left panel presents low-mass galaxies ($\log(M_{\mathrm{tot}}/M_{\odot}) < 10.6$),  while the right panel shows high-mass systems  ($\log(M_{\mathrm{tot}}/M_{\odot}) \geq 10.6$). The increase in $I_{\mathrm{N}}$ with $d_{\mathrm{n}}$ indicates that the statistical coupling between mass composition and colour becomes progressively stronger away from nodes. The effect is most pronounced for low-mass galaxies, while high-mass systems show a substantially weaker environmental dependence.

For low-mass galaxies, the environmental trend seen in the full sample is reproduced with comparable or even enhanced contrast. At fixed filament distance, $I_{\mathrm{N}}$ increases strongly with node distance, reaching values $I_{\mathrm{N}} \gtrsim 0.28$ in regions that are filament-dominated but node-remote. This confirms that low-mass systems are particularly sensitive to the competing influences of nodal processing and filamentary gas regulation. Their shallower potential wells render them more vulnerable to tidal stripping and ram-pressure effects near nodes, which efficiently disrupt the intrinsic mass-colour coupling and suppress $I_{\mathrm{N}}$.

In contrast, for high-mass galaxies the $I_{\mathrm{N}}$ map is noticeably flatter. Although modest environmental variation persists, the amplitude of change with node distance is significantly reduced. Even near nodes, $I_{\mathrm{N}}$ does not decline as sharply as in the low-mass case. This behaviour shows that massive galaxies maintain a comparatively stable relationship between mass composition and colour across environments in the EAGLE simulation. Their deeper gravitational potentials and more efficient internal feedback processes reduce susceptibility to environmental decoupling, so that external dynamical processing near nodes does not substantially erase the structural imprint of baryonic composition.

\autoref{fig:nmi_1d} reveals a distinct environmental modulation of the mass-colour coupling. In the full sample (left panel), the normalised mutual information $I_{\mathrm{N}}$ increases steadily with node distance up to $d_{\mathrm{n}} \sim 2~\mathrm{Mpc}$ and then approaches a plateau at larger separations. This behaviour reveals a characteristic scale at which environmental processes associated with nodes in the EAGLE simulation most strongly influence the statistical link between mass composition and galaxy colour. Beyond this scale, the coupling becomes stable. The mass-segregated trends (right panel) clarify the origin of this behaviour. High-mass galaxies exhibit only a weak dependence of $I_{\mathrm{N}}$ on $d_{\mathrm{n}}$, remaining nearly uniform across the explored range, indicating that their mass-colour connection is relatively insensitive to nodal proximity. In contrast, low-mass galaxies show a pronounced rise in $I_{\mathrm{N}}$ with increasing $d_{\mathrm{n}}$, extending to $d_{\mathrm{n}} \sim 4~\mathrm{Mpc}$ before flattening. This stronger and more extended environmental dependence is consistent with the greater response of low-mass systems to the environmental processes operating near nodes within the EAGLE framework, and indicates that the decoupling of mass structure and colour is primarily driven by cluster-scale effects acting on shallow potential wells.

Taken together, \autoref{fig:fig4}- \autoref{fig:nmi_1d} demonstrate that the environmental dependence of the mass-colour coupling is intrinsically mass dependent. Nodal environments preferentially disrupt the structural imprint of baryonic composition on colour in low-mass galaxies, while massive systems evolve in a manner more strongly regulated by internal processes. Filaments, on the other hand, preserve and enhance the intrinsic mass-colour linkage by modulating gas supply in a gradual and mass-dependent fashion. The combined behaviour highlights the interplay between internal binding energy and large-scale tidal geometry in shaping the information content shared between galaxy mass structure and evolutionary state.

\begin{figure}
\centering
\includegraphics[width=1.5 \columnwidth/2]{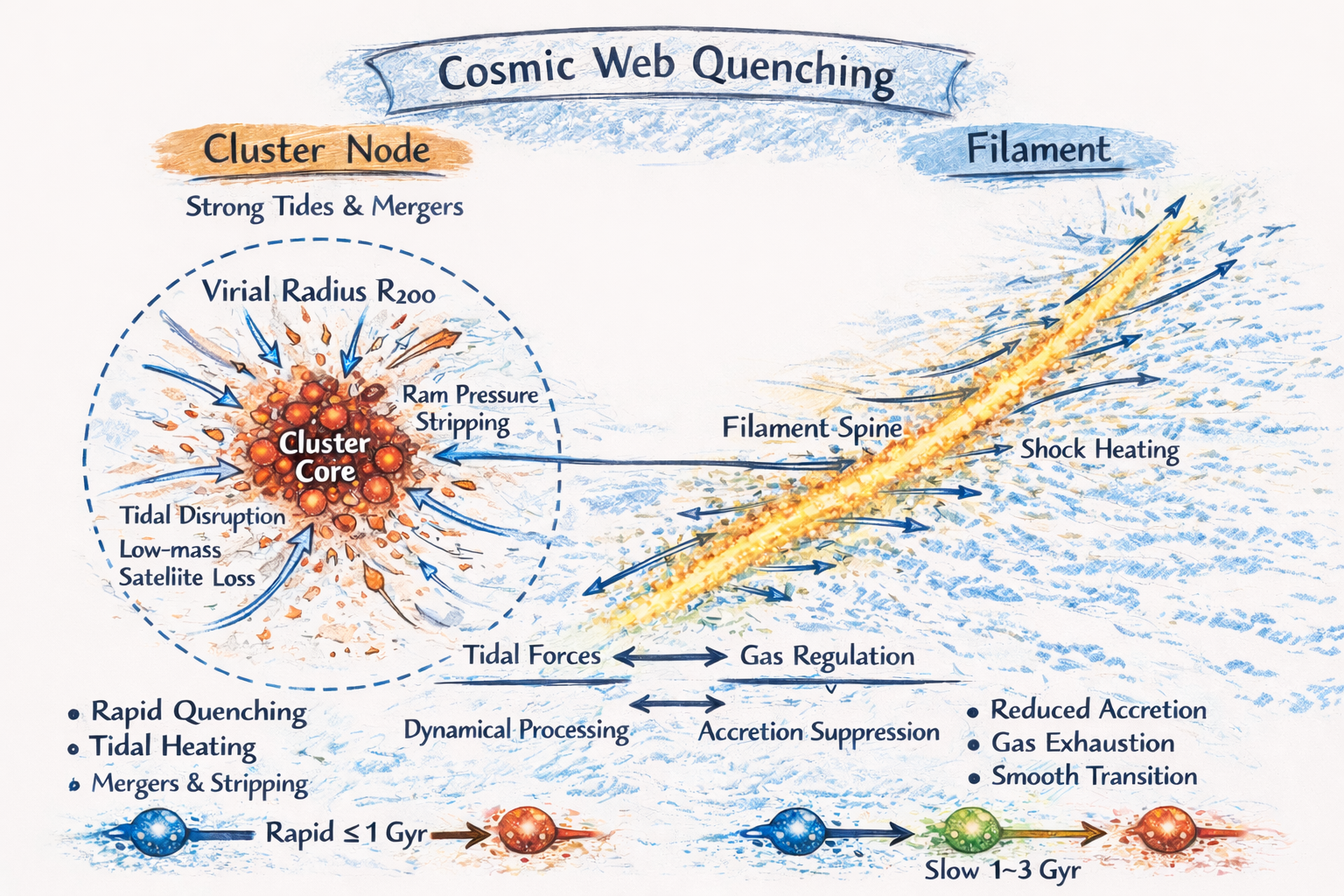}
\caption{This schematic illustration shows quenching mechanisms in
  nodes and filaments.  Galaxies in cluster environment experience
  strong tidal fields, ram-pressure stripping, and enhanced merger
  activity within the virial region ($\sim R_{200}$), leading to rapid
  or externally driven quenching and partial decoupling between
  internal mass structure and colour.  In filaments, gas and galaxies
  flow along the filament spine toward nodes. Quenching proceeds
  gradually through suppression of gas accretion (cosmic web
  starvation), preserving a stronger coupling between mass composition
  and colour.}
\label{fig:schematic_quenching}
\end{figure}

\autoref{fig:schematic_quenching} encapsulates the central physical picture emerging from our analysis: nodes and filaments regulate galaxy evolution through fundamentally different quenching channels. In the EAGLE simulation, cluster-scale nodes (left) correspond to virialised regions typically associated with massive halos and clusters. As galaxies approach these environments, several quenching mechanisms such as ram-pressure stripping, tidal interactions, harassment, and shock heating of infalling gas become increasingly effective. These processes efficiently remove or heat the cold gas reservoirs of galaxies, thereby suppressing star formation and increasing the red fraction. This often disrupts the intrinsic link between a galaxy’s internal mass structure and its star formation activity. In contrast, filament environments (right) are governed by anisotropic inflows along the spine, where gas and galaxies stream coherently toward nodes. Within filamentary environments, the implemented gas accretion and galaxy growth processes generate more gradual trends in galaxy properties across the filament cross-section, allowing the colour transformation to remain more tightly coupled to the galaxy’s internal baryonic composition. This highlights the scale-dependent and mechanism-dependent nature of cosmic web quenching where nodes and filaments give rise to distinct environmental signatures, reflecting the different dynamical and gas-regulation processes operating within these environments in the EAGLE simulation.

\section{Conclusions}
\label{sec:conclusions}

In this work, we have investigated how the two principal structural components of the cosmic web i.e nodes and filaments jointly and independently regulate galaxy evolution. By combining relative colour fractions with an information-theoretic analysis of the coupling between galaxy colour and mass composition, we have uncovered a coherent, scale-dependent picture of cosmic web quenching.

We identify two characteristic environmental scales associated with nodes and filaments, indicating that these components leave distinct signatures on galaxy populations. Red galaxies preferentially inhabit regions close to filament spines and nodes, while blue galaxies dominate at larger distances. The characteristic crossover scales $\sim 0.75~\mathrm{Mpc}$ for filaments and $\sim 2.5~\mathrm{Mpc}$ for nodes are physically meaningful. The node crossover scale is comparable to typical virial radii ($R_{200}$) of group- and cluster-scale halos, indicating that quenching becomes efficient once galaxies enter dynamically bound regions. Within $\sim R_{200}$, environmental processes such as ram-pressure stripping, tidal heating, and mergers operate on timescales comparable to or shorter than the dynamical time ($t_{\mathrm{dyn}} \leq 1~\mathrm{Gyr}$), driving accelerated transformation toward the red sequence. This naturally explains the monotonic rise in the red fraction with decreasing $d_{\mathrm{n}}$ (\autoref{fig:fig1}). Interestingly, we observe a turnover in the red fraction at $d_{\mathrm{f}} \lesssim 0.5~\mathrm{Mpc}$. The turnover in the red fraction at smaller $d_{\mathrm{n}}$ is associated with the competing interplay between environmental quenching and dynamical selection effects in the innermost cluster regions. Galaxies are particularly vulnerable to tidal disruption and merging in high-density cores. Consequently, the very central regions may become increasingly dominated by massive galaxies and remnants of previous mergers, reducing the number of surviving red systems. In addition, recently accreted galaxies undergoing rapid environmental transformation may temporarily populate intermediate colour states, thereby lowering the instantaneous red fraction in the very central regions. Therefore, the decline in the red fraction at very small node distances is consistent with a population mix produced by the environmental processes operating within cluster environments in EAGLE. In contrast, the filament crossover scale traces the transverse extent of dynamically influential filament cores. Here quenching is more gradual and regulated by the suppression or modulation of cold gas accretion. Gas depletion over $\sim 1$-$3~\mathrm{Gyr}$ timescales naturally produces a smooth migration from the blue cloud through the green valley to the red sequence. The monotonic dependence of red fraction on $d_{\mathrm{f}}$ and the absence of a sharp turnover indicate that filaments act as extended regulators of gas supply rather than sites of impulsive stripping. The persistence of the observed environmental trends even within narrow stellar-mass and halo-mass intervals indicates that the characteristic scales identified in this work cannot be explained solely by mass segregation across the cosmic web. Instead, these results point to an intrinsic geometric influence of large-scale structure on galaxy populations. This reinforces the view that nodes and filaments imprint distinct environmental signatures on galaxy evolution beyond the effects of halo mass alone. Overall, these trends are consistent with the physical picture illustrated in \autoref{fig:schematic_quenching}.

The normalised mutual information analysis provides a complementary and deeper perspective. We find that the statistical coupling between dominant mass component and galaxy colour strengthens with increasing node distance at fixed filament proximity. Near nodes, intense dynamical processing partially erases the intrinsic link between baryonic composition and star formation activity, reducing the predictive power of internal mass structure. Far from nodes, within filament-dominated regions, colour remains more tightly coupled to mass composition, implying that filament-driven quenching is structurally mediated. The mass dependence of this behaviour is particularly revealing. Low-mass galaxies exhibit strong environmental modulation in both relative fractions and mutual information, consistent with their vulnerability to external tidal forces and gas stripping within $\sim R_{200}$. Massive galaxies, by contrast, display flatter environmental trends, indicating that deeper gravitational potentials and internal feedback processes buffer them against environmental decoupling. This dichotomy highlights the interplay between gravitational binding energy and large-scale tidal geometry in shaping galaxy evolution. 

These results naturally align with the framework of anisotropic assembly bias. Filaments correspond to regions of one-dimensional collapse, where coherent matter inflow modulates halo growth and gas accretion histories at fixed mass. Nodes represent sites of multi-axial collapse and nonlinear dynamical evolution, where mergers and tidal heating accelerate structural transformation. The observed environmental gradients and the spatial variation of mass-colour coupling together suggest that galaxy properties retain memory of their anisotropic assembly pathways, but that this memory can be partially erased within virialised cluster environments. Cosmic web quenching therefore emerges as a hierarchical and scale-dependent phenomenon. Filaments regulate galaxies over megaparsec scales by modulating gas accretion along coherent tidal directions, while nodes impose rapid dynamical processing within virial radii over dynamical timescales. The combined action of these environments shapes not only the demographic distribution of red and blue galaxies, but also the information content linking internal mass structure to evolutionary state.

In summary, our results demonstrate that galaxy quenching cannot be understood solely in terms of local density or halo mass. The geometry of the cosmic web, encoded through distances to filament spines and nodes plays a fundamental and physically distinct role. By explicitly separating these geometric effects and quantifying their impact through both population statistics and information theory, we provide a unified and physically grounded picture of how large-scale structure sculpts galaxy evolution across cosmic time.

Our results are broadly consistent with earlier studies that reported enhanced red fractions and suppressed star formation in proximity to filaments and nodes, even after controlling for local density \citep{kuutma17, malavasi17, laigle18, kraljic18, bonjean20, pandey20, maret24, guang25}. Our results indicate that the physical processes implemented in EAGLE manifest themselves differently in nodal and filamentary environments. The systematic increase of passive galaxies toward filament spines and cluster-scale nodes therefore reinforces the emerging view that cosmic web geometry encodes environmental information beyond simple density contrasts. At the same time, our analysis advances this picture in two important ways. First, by jointly analysing distances to both nodes and filaments, we identify distinct statistical signatures associated with nodal and filamentary environments, reflecting the different ways in which the physical processes implemented in EAGLE manifest across the cosmic web. Second, by employing an information-theoretic framework, we show that nodal environments partially erase the intrinsic coupling between mass composition and colour, whereas filamentary environments preserve and even enhance it. This reveals not merely where galaxy populations change across the cosmic web, but also how the statistical imprint of assembly history is encoded in observable galaxy properties. Together, these results refine and deepen the geometric paradigm of environmental quenching, providing a more physically resolved understanding of how large-scale structure shapes galaxy evolution.

An important aspect of this work is that many of the environmental signatures identified in the EAGLE simulation can be directly compared with observations. The enhanced abundance of red galaxies and the corresponding decline of blue populations toward filament spines and nodes are qualitatively consistent with trends reported in SDSS, GAMA, VIPERS, COSMOS, and other surveys \citep{malavasi17, laigle18, kraljic18, kuutma17, pandey20, hoosain24, zarattini25}. These observational studies have already established that galaxy colours, star-formation activity, and gas content exhibit systematic gradients with cosmic web environment, supporting the broad physical picture implemented in the simulation. At the same time, several of the statistics introduced here constitute more specific predictions. In particular, the characteristic transition scales associated with nodes and filaments, the differential behaviour of low- and high-mass galaxies, and the spatial variation of the mass-colour coupling quantified through normalised mutual information provide new diagnostics that can be tested observationally. Such measurements require reliable reconstruction of the three-dimensional cosmic web and well-characterised galaxy properties, making spectroscopic surveys particularly valuable. Existing datasets such as SDSS and GAMA already offer promising opportunities for nearby-Universe tests, while ongoing and forthcoming surveys including DESI, Euclid, and Rubin LSST will enable substantially more precise measurements across larger cosmological volumes and a wider redshift range. Agreement between these observations and the signatures identified here would lend support to the physical processes implemented in current hydrodynamical simulations, whereas significant discrepancies could point to missing or incomplete aspects of galaxy formation physics.

Looking forward, these results offer several promising avenues for both theoretical refinement and observational validation. On the theoretical side, extending this analysis across redshift would allow direct examination of how virial-scale processing and filamentary regulation evolve with cosmic time, particularly during the peak epoch of star formation when gas accretion rates were higher and cluster environments less dynamically mature. Incorporating halo accretion histories and tidal tensor eigenvalue analyses would further clarify the connection between anisotropic assembly and baryonic transformation. Observationally, forthcoming wide-field spectroscopic surveys such as Euclid \citep{euclid}, DESI \citep{desi}, and LSST \citep{lsst} will enable precise reconstruction of the three-dimensional cosmic web over large volumes, allowing measurements of colour fractions and gas content gradients relative to filament spines and cluster nodes. Combining such surveys with HI and molecular gas observations will be particularly powerful in testing the predicted distinction between rapid, virial-scale quenching in nodes and gradual, accretion-regulated quenching in filaments. Ultimately, bridging simulations and observations in this geometrically explicit framework will provide a decisive test of whether anisotropic assembly and cosmic web geometry constitute a fundamental driver of galaxy evolution.

\section{Acknowledgements}
The authors sincerely thank an anonymous reviewer for insightful comments and suggestions that helped to improve the draft. BP gratefully acknowledges financial support from the Government of India under the project  project ANRF/ARG/2025/000535/PS. BP also thanks IUCAA, Pune for support received through the Associateship Programme. AD acknowledges Harish-Chandra Research Institute, Allahabad for support provided through a postdoctoral fellowship. The authors extend their gratitude to the Virgo Consortium for providing public access to their simulation data. The EAGLE simulations were performed using the DiRAC-2 facility at Durham, managed by the ICC, and the PRACE facility Curie based in France at TGCC, CEA, Bruy\`{e}res-le-Ch\^{a}tel.

\section{Data availability}
The data from the EAGLE simulations employed in this work are publicly available through the EAGLE database (\url{https://icc.dur.ac.uk/Eagle/database.php}). Any supplementary data products produced in this study are available from the authors upon reasonable request.

\bibliographystyle{JHEP}
\bibliography{refs.bib}

\label{lastpage}
\end{document}